\documentclass[12pt]{article}

\usepackage{newtxtext,newtxmath}
\usepackage{graphicx}
\usepackage{lineno}
\usepackage[utf8]{inputenc}
\usepackage[T1]{fontenc}
\usepackage{color}
\usepackage{ragged2e}
\usepackage{threeparttable}
\usepackage{booktabs}
\usepackage{booktabs, makecell, multirow, tabularx}
\usepackage{float}  %设置图片浮动位置的宏包
\usepackage{subfigure}  %插入多图时用子图显示的宏包

\usepackage[letterpaper,margin=1in]{geometry}

\renewenvironment{abstract}
	{\quotation}
	{\endquotation}

\date{}

\makeatletter
\renewcommand{\fnum@figure}{\textbf{Figure \thefigure}}
\renewcommand{\fnum@table}{\textbf{Table \thetable}}
\makeatother

\usepackage{scicite}

\usepackage{url}

\def\scititle{
	High-key-rate Fully-Passive Quantum Access Network with Thermal Source
}
\title{\bfseries \boldmath \scititle}

\author{
	Hanwen Yin$^{1\dagger}$,
	Bangde Zhu$^{1,\dagger}$,
    Peng Huang$^{1,2,3,\ast}$,
    Tao Wang$^{1,2,3}$,\and
    Xueqin Jiang$^{3,4}$,
    Yankai Xu$^{1}$,
    Guihua Zeng$^{1,2,3,5,\ast}$\and
        \small$^{1}$State Key Laboratory of Photonics and Communications \& Center for Quantum Sensing and Information Processing, Shanghai Jiao Tong University, Shanghai, 200240, China, \and
        \small Shanghai Jiao Tong University, Shanghai 200240, China \and
        \small$^{2}$Shanghai Research Center for Quantum Sciences, Shanghai 201315, China \and
        \small$^{3}$Hefei National Laboratory, Hefei 230088, China \and
        \small$^{4}$College of Information Science and Technology, Donghua University, Shanghai, 201620, China \and
        \small$^{5}$Shanghai XunTai Quantech Co., Ltd, Shanghai, 200241, China \and
	\small$^\ast$Corresponding author. Email: huang.peng@sjtu.edu.cn, ghzeng@sjtu.edu.cn\and
	\small$^\dagger$These authors contributed equally to this work.
}

\begin{document} 
% Insert the title and author list
\maketitle

% Abstract, in bold
% There are strict length limits, and not all formats have abstracts.
% Consult the journal instructions to authors for details.
% Do not cite any references in the abstract.
\begin{abstract} \bfseries \boldmath
% Start with one or two sentences of background
To accommodate classical communication systems with progressively increasing transmission rates, quantum access networks (QAN) have undergone systematic and protocol-level optimizations in recent years, where quantum passive optical network (QPON) architectures are gaining significant attention due to their simple structure. It is challenging for the previous QAN based on active protocols or Stokes operator coding protocols to achieve high-speed linear modulation with high extinction ratio and stability under practical conditions. In this work, we propose and experimentally demonstrate a downstream fully passive quantum access network protocol using passive state preparation (PSP) with free-space and single-mode fiber hybrid channels, and the final key generation rate is up to a record-breaking 19.48 Mbps per quantum network unit. The proposed PSP-QPON scheme extends the scope of PSP-CVQKD from point-to-point to point-to-multi-point networks, which enables high-key-rate, high-stability, and low-resource-consumption implementation. Moreover, the network channel in this experiment is fully compatible with access networks in classical optical communications, which allows integration with existing optical infrastructure without the need for additional modifications, providing a promising solution for local area network quantum access network at home or a mobile terminal.  
\end{abstract}

\section*{INTRODUCTION}
% The first paragraph of any Science paper does NOT have a heading
% Nor is it indented
\noindent
Quantum key distribution (QKD) provides an information-theoretic secure communication method\cite{ref2,ref8,ref9,ref10,ref11,ref12,ref13,ref14}. Its development is advancing from point-to-point systems toward networking, achieving multi-scenario deployments ranging from metropolitan area networks\cite{QN_1,QN_2,QN_3,QN_4} and wide area networks to space-to-ground links\cite{fs-dv1,fs-dv2,fs-dv3,fs-dv4}. As the "last mile" of QKD networks, the quantum access network (QAN) is of significant importance for the large-scale application of QKD, and relevant research has attracted considerable attention. As early as 1997, a multi-user QKD scheme based on a downstream passive optical fiber network was proposed \cite{QAN_1997}, and subsequent other QAN schemes have gradually emerged \cite{QAN_2013}. Although progress has been made in cost control and other aspects, problems such as a limited secure key rate still exist. While continuous-variable (CV) QKD can achieve a high key rate, high integration, and low-cost communication under relatively low transmission loss, it fundamentally possesses strong resistance to background light noise by virtue of quantum coherent detection. The development and improvement of QAN schemes based on CVQKD have effectively broken through the technical bottlenecks of traditional quantum access networks, providing a feasible path for the large-scale popularization and application of QKD, and have become an important research focus and development direction in the current field of QANs \cite{QAN_46node,QAN_10Gb,huang2020experimental,QAN_roundtrip}.

To accommodate classical communication systems with progressively increasing transmission rates, such as Light Fidelity (LiFi) and multi-band modulation systems, quantum access networks have undergone systematic and protocol-level optimizations in recent years. Consequently, related research has achieved increasingly higher achievable key rates \cite{Huang:20,Qi:24,QAN_AA_2024,QAN_16node}. However, the previous CVQKD protocols all relied on active protocols such as Gaussian-modulated coherent states (GMCS) or Stokes operator coding protocols. For the active CVQKD protocol, the modulated quantum states are always prepared actively by using amplitude and phase modulators, and achieving high-speed linear modulation with a high extinction ratio and stability under practical conditions is challenging, especially in integration design. Moreover, in an active CVQKD protocol, high-speed on-chip modulators significantly increase costs, manufacturing time, and implementation complexity \cite{psp2}, which can not fully meet the requirements of QAN.

To overcome the challenges of traditional GMCS CVQKD with active modulation, B. Qi et al. proposed the passive-state-preparation (PSP) CVQKD protocol using a thermal source in 2018 \cite{psp1}. Since the avoidance of quantum coherent sources and active modulations and the feasibility of secure key generation over metro-area distances \cite{psp2}, the PSP CVQKD protocol has shown great potential in QKD networks \cite{psp3,psp4,psp6}. Recently, the implementation of a PSP CVQKD experiment with a 1.09 Gbps key rate \cite{psp5} and an all-day field trial \cite{psp7} further demonstrated the potential of the PSP CVQKD protocol in achieving high-key-rate systems. Due to its high resistance to background noise, impressive bit rate capabilities, and low power consumption, the PSP CVQKD system is particularly well-suited for secure communication onboard UVs. However, the previous research on PSP CVQKD has focused solely on the PTP protocol to explore its application scenarios. A critical gap remains in PSP QAN regarding the extension of the PSP CVQKD protocol from PTP systems to scalable networks, which is a key prerequisite for large-scale deployment.

In this article, we propose a PSP quantum passive optical network (PSP-QPON), which first extends PSP-CVQKD from PTP architectures to PTMP networks. Focusing on the downstream link, we conduct a dedicated experiment over a custom-built hybrid channel: consisting of a 5 km single-mode fiber and a free-space segment with an average attenuation of -4 dB, the hybrid channel achieves an overall average attenuation of -10.96 dB. For hybrid channel adaptability, we optimize the system’s noise suppression and signal compensation algorithms, ensuring stable operation even under uneven attenuation conditions. In terms of scalability-rate balance, the passive optical splitting design of PSP-QPON avoids signal degradation caused by active components, supporting efficient multi-user access. Notably, in a 1-to-4 PTMP network configuration, our proposed PSP-QPON delivers a secure key rate of 19.48 Mbps per user. These results validate our scheme's ability to break the PTP bottleneck of traditional PSP-CVQKD, providing a high-rate, scalable and cost-effective solution for constructing practical QKD networks. This lays a solid foundation for the next generation of large-scale secure optical communication infrastructures.
\section*{RESULTS}

\subsection*{Network architecture and protocol}

\begin{figure*}[htbp]
        \centering
        \includegraphics[width=1\linewidth]{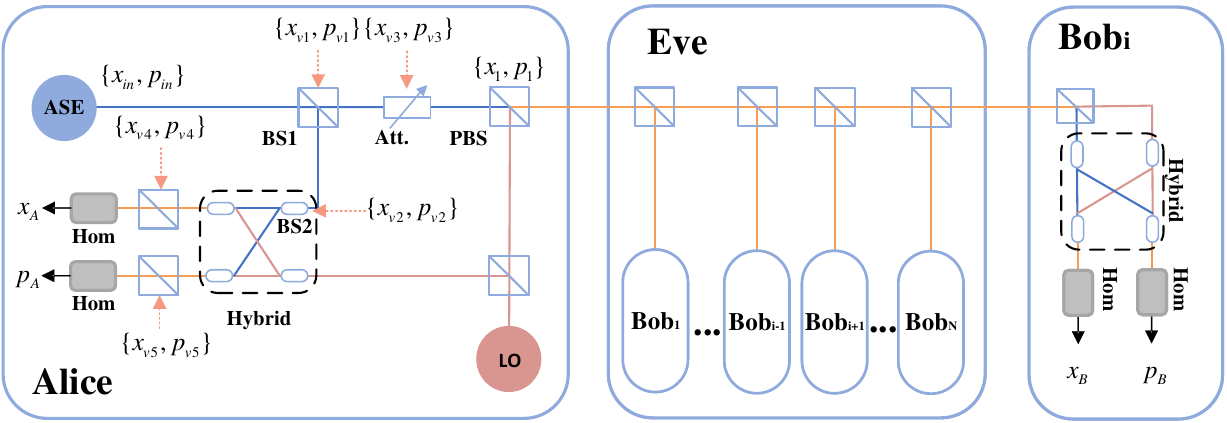}
        \caption{\textbf{Optical layout of the PSP-QPON using a thermal source.} Alice prepared quantum states and sent them to each Bob through a passive splitter, while Eve may eavesdrop on information with the assistance of any number of Bobs. The red, blue, and yellow fibers denote LO, quantum signal, and polarization multiplexing signal, respectively. ASE, amplified spontaneous emission; LO, local oscillator; BS, beam splitter; PBS, polarization beam splitter; VOA, variable optical attenuator;  Hom, homodyne detection.} 
        \label{fig: protocol}
\end{figure*}

\textbf{QPON architecture} The network architecture of PSP-QPON is shown in ~Figure~\ref{fig: protocol}, which integrates the high key rate of the PSP protocol and the high capacity of passive networks. The entire network architecture is divided into quantum line terminal (QLT) and quantum network unit (QNU), and all nodes are connected through passive optical components such as single-mode fibers and 1:N passive splitters.

\textit{QLT (Alice):} As the source of passively prepared quantum states, the QLT does not require active modulation. QLT splits the output signal of a thermal source into two spatial modes using a beam splitter (BS). One mode $\{x_\text{1},p_\text{1}\}$ is sent to Bob after passing through a VOA, while the other mode $\{x_\text{A},p_\text{A}\}$ is measured locally by QLT using heterodyne detection. Since QLT cannot acquire the information about the quantum signal directly, she must obtain her optimal estimation of the outgoing mode \cite{psp1} from her local measurement results. And the uncertainties derived from QLT’s estimation of the outgoing mode can be described as additive excess noise, i.e., the PSP noise (see Supplementary Information, Note 1). The modulation variance $V_\text{A}$ depends on the source intensity, the splitting ratio of the first BS, and the optical attenuation.

\textit{QNU (Bob):} Each QNU node performs heterodyne detection on the received optical mode to obtain the quadratures $x_\text{B}$ and $p_\text{B}$. To extract more information from noisy data, the QNU node also incorporates other essential optical instruments. QLT and QNUs further estimate the parameters using a subset of their correlation data through an unreliable classical channel. The mutual information between Alice and the various Bobs changes as their received mode evolve through different channels, enabling each Bob to extract a secure key from the broadcast information.

\textbf{QPON protocol} In the PSP-QPON protocol, all QNUs can obtain unique measurement results in each protocol round. This uniqueness stems from the independent quantum noise each QNU experiences. Through reverse information reconciliation, the QLT can simultaneously generate independent keys with each QNU; after privacy amplification, these keys become fully decoupled from Eve and other QNUs, eliminating information leakage risks (see Supplementary Information, Note 2).

From the QNU's view, he cannot distinguish the optical mode prepared in PSP protocol and the traditional GMCS protocol, and the same goes for Eve. Therefore, the security of the PSP-QPON protocol is guaranteed by its equivalence to the traditional QPON protocol, and the asymptotic secure key rate calculation between QLT $A$ and QNU $B_i$ under collective attacks is
\begin{equation}
    K_i=f(1-FER)\left[\beta I(A:B_i)-\max{\left\{ I(B_i:\bar{B}_i),\chi_{B_iE}\right\}}\right]
\end{equation}
where $f$ is the equivalent repetition rate, i.e. the bandwidth of the homodyne detector; $FER$ is the frame error rate of the reconciliation step; $\beta$ is the reconciliation efficiency, $I(A: B_i)$ is the classical mutual information between QLT and QNU $i$, $I(B_i:\bar{B}_i)$ is the classical mutual information between QNU $i$ and the other QNUs in the network, $\chi_{B_iE}$ is the Holevo information between QNU $i$ and Eve. In this formulation, the key point is that we need to remove not only the information that Eve may have stolen by attacking the channel between QLT and QNU $i$ but also any residual information that the other QNUs may have about QNU $i$, so the key is secret and also independent for each QNU.
Here, the other QNUs in the network are considered to be trustworthy, meaning that they will not collaborate with Eve, and she cannot acquire any information from them. 

Under standard experimental conditions, the contribution of $\chi_{B_iE}$ typically exceeds that of $I(B_i:\bar{B}_i)$ \cite{QAN_16node}. This relationship simplifies the secure key rate to the point-to-point protocol
\begin{equation}
    K_i=f(1-FER)\left[\beta I(A:B_i)-\chi_{B_iE}\right].
    \label{equ: skr}
\end{equation}
Therefore, the correlations among QNUs do not affect the protocol's performance in this condition. The PSP-QPON enables a fully passive network with a high bit rate and large QNU capacity without active modulation. It breaks through the limitations of traditional quantum access networks that often trade off between key rate and QNU scale, and realizes the optimization of security performance and network expansion capability through the inherent characteristics of passive quantum state preparation and distribution

\subsection*{Experimental setup}

\begin{figure*}[htbp]
        \centering
        \includegraphics[width=1\linewidth]{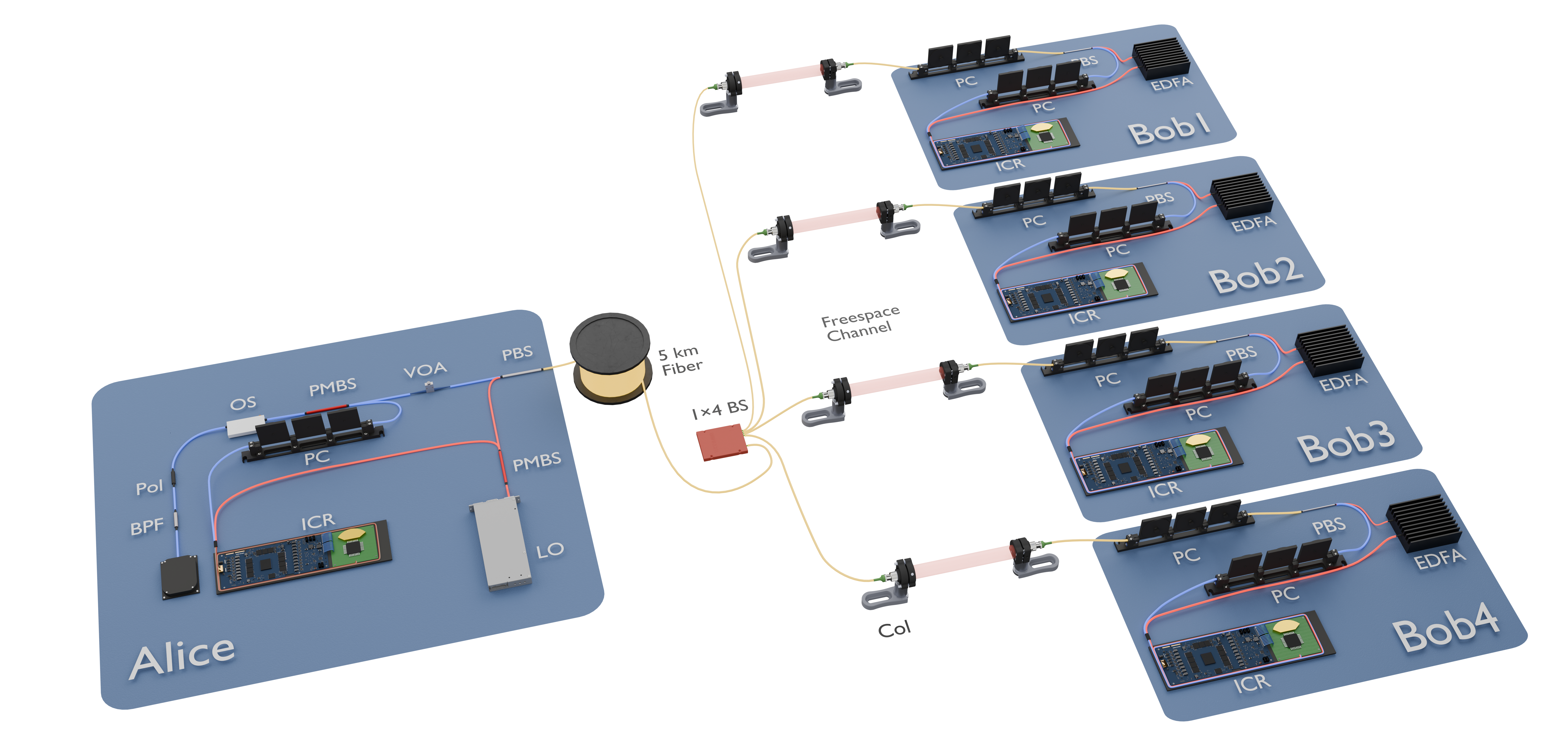}
        \caption{\textbf{The experimental setup of PSP-QPON.} The red, blue, and yellow fibers denote LO, quantum signal and polarization multiplexing signal, respectively. ASE, amplified spontaneous emission; LO, local oscillator; BPF, bandpass filter; Pol, polarizer; PMBS, polarization maintain beam splitter; PBS, polarization beam spliter; VOA, variable optical attenuator; Col, collimator; EDFA, Er-Doped fiber amplifier; PC, polarization controller; OSW, optical switch; ICR, integrated coherent receiver.} 
        \label{fig: setup}
\end{figure*}

The experimental setup of the PSP-QPON is shown ~Figure~\ref{fig: setup}. To present a free-space local area network (LAN), the experimental channel was designed as a free-space single-mode fiber hybrid channel, where the sending light first passes through a 5 km fiber channel to simulate a LAN fiber to the home. After that, a 1-to-4 beam splitters consisted by three 50:50 beam splitter are employed to send it to each of the four Bobs over a 1 m indoor free-space channel constructed using two collimators with an average -4.3 dB attenuation for each Bob to simulate a Li-Fi or a short-range free-space channel condition. Once each Bob has measured the states sent from Alice and achieved the raw data, every two parties post-process their raw data to generate a secret key, via parameter estimation, information reconciliation, error correction, and privacy amplification.

On Alice's side, a 0.4 nm optical bandpass filter with 1550.12 nm central wavelength and a polarizer with 30 dB polarization extinction ratio is placed after the ASE source to get a single-mode thermal state (see Supplementary Information, Note 3). A continuous-wave laser source centered at 1550.12 nm with 0.1 kHz line width is employed as local oscillator (LO), and the launching power is set to 10 dBm. A PMBS with a 99:1 splitting ratio is used to split the LO, where the weaker mode is sent to Bob to reduce the leakage noise. A thermal source is used as the signal, and it is split by a BS with a 99:1 splitting ratio to reduce the excess noise due to passive state preparation. The weaker mode is modulated with a VOA to get an appropriate equivalent modulation variance and sent to Bob with LO by polarization multiplexing through PBS, while the other mode is measured locally by Alice with heterodyne detection. Here, an ICR is used to implement the heterodyne detection. Since the signal optical path is not polarization-maintained, a PC is placed in front of the signal optical path of ICR to ensure the correct polarization state of the input, which ultimately manifests as detection efficiency for the QNU.

On Bob's side, the polarization state of the signal is modulated by the PC, and then the optical signal is polarization-demultiplexed by a PBS with a high extinction ratio. A 99:1 BS is placed after the quantum signal path to split 1\% of the signal, which helps us monitor and control the polarization state of the received signal to reduce leakage noise. To reach the shot noise limit, a polarization-maintained EDFA with a low noise figure is employed to amplify the LO signal. To acquire real-time shot noise, a polarization-maintained OSW derived by an arbitrary function generator (AFG) is placed on the signal path after the PBS for real-time shot-noise unit (SNU) calibration. The OSW can provide a maximum extinction ratio of 24 dB, and the signal power is suppressed from about 4 SNU to about 0.04 SNU. The polarization state of the signal light is adjusted using a PC on Alice's side, and an ICR is employed to implement heterodyne detection.

The data are acquired synchronously by an oscilloscope. A 20 GHz bandwidth real-time oscilloscope is employed to over-sample the outputs of Alice's and Bob's ICR with a 4 GHz bandwidth. A detection window of 100 microseconds is chosen, and the output signal of the ICR is oversampled at a sample rate of 40 GHz. After Alice and Bob share correlated raw data, the DSP algorithms, including digital filter, frame synchronization, and phase compensation, are employed to directly correlate the sampling points between Alice and Bob and decrease the excess noise. Specifically, based on the concept of continuous temporal mode, a machine learning based signal demodulation algorithm is proposed, and the SNR of the signal has been significantly improved. In the parameter estimation procedure, the covariance matrix $\gamma_{AB_1...B_N}$ is estimated, as shown in~Figure~\ref{fig: mutu}a, based on which the SKR between A and each Bi can be calculated as in~Equation~\ref{equ: skr}. Each 100 consecutive data frames is spliced into a final data frame. To get the real-time shot noise, an optical switch is employed to randomly switch between the measurement signal and shot noise. For each data frame, a total of 10 sets of shot noise are counted before and after it to minimize statistical error in shot noise estimation.

\subsection*{Experimental results}

The overview of key experimental parameters of the four QNUs PSP-QPON is shown in~Table~\ref{tab: para}. The reconciliation efficiency is 96\%, and the equivalent repetition frequency is 4 GHz determined by the detector bandwidth. Theoretically, a QKD system possesses an optimal modulation variance that yields the best performance under current conditions. However, in practical experiments, increasing the modulation variance may lead to higher excess noise levels, attributable to the residual phase noise. Therefore, in this experiment, we set the modulation variance to 4.28 SNU. The detection efficiency is 0.56. The link losses for the four Bobs are -10.77 dB, -11.21 dB, -10.94 dB, and -11.10 dB, and the detailed transmittance of the hybrid channel is shown in~Figure~\ref{fig: mutu}c (see Supplementary Information, Note 4).

\begin{figure*}[htbp]
        \centering
        \includegraphics[width=0.72\linewidth]{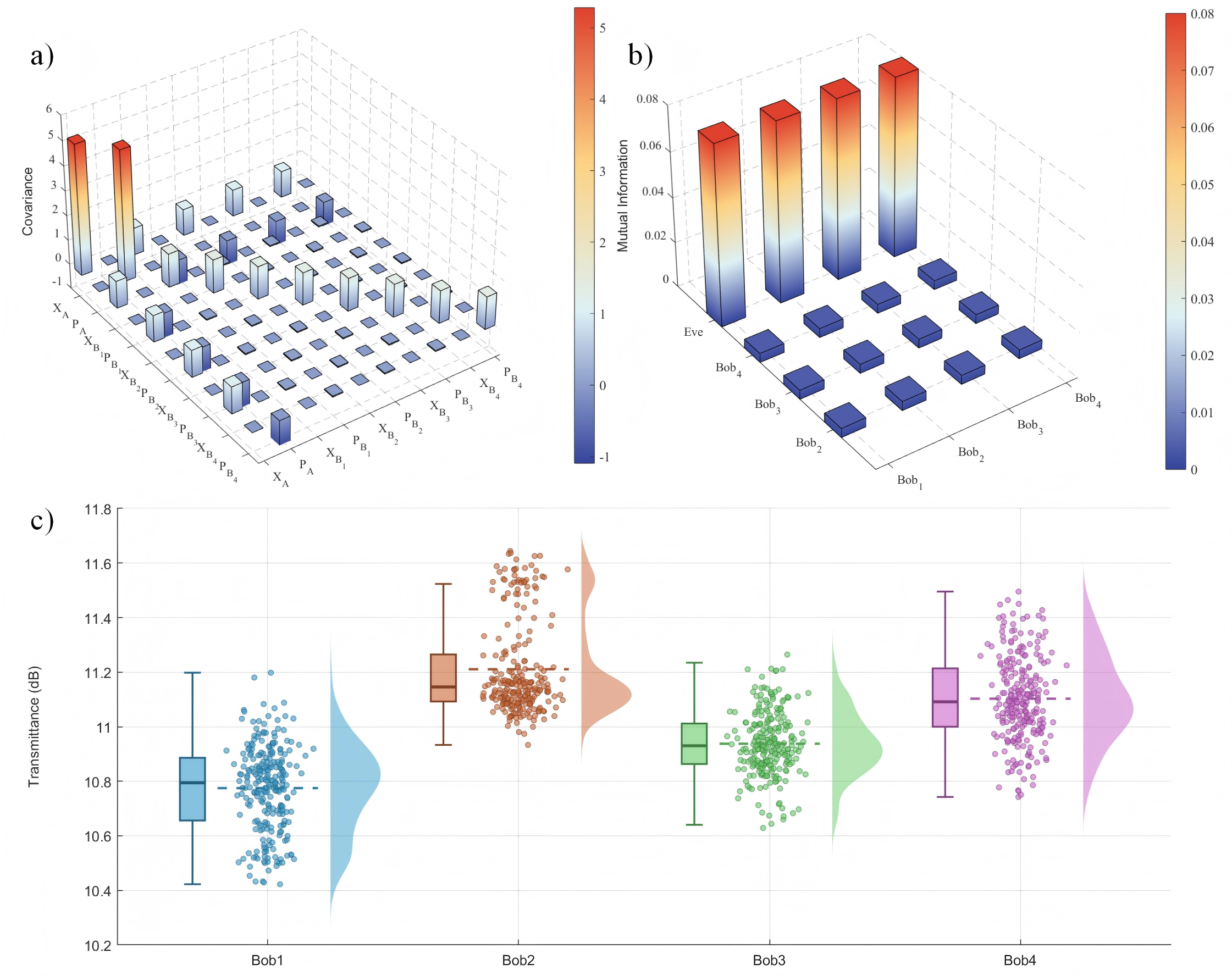}
        \caption{\textbf{The mutual information, covariance, and transmittance.} a) The covariance matrix of the system. b) The classical mutual information between each node and the Holevo bound between Eve and each receiver mode. c) The transmittance for each node.}
        \label{fig: mutu}
\end{figure*}

The correlation between different Bobs can be defined by classical mutual information, denoted as $I_{B_iB_j}$. In the 4-node hybrid channel, the mutual information between different Bobs is shown in~Figure~\ref{fig: mutu}b. The maximum mutual information is 0.0039 bit/pulse, which is far lower than the minimum Holevo bound between Eve and Bob where $\chi_{BE}^{min}=0.11$ bit/pulse. Therefore, the calculation of SKR degenerates into the basic point-to-point condition in~Equation~\ref{equ: skr} (see Supplementary Information, Note 5).

The~Figure~\ref{fig: skr} shows estimated excess noise and SKR for each QNU. Here, we acquired 1200 data frames with block size of $4 \times 10^5$. Each test round is estimated by 25 consecutive frames, so the block size corresponding to the parameter estimation results of each point in the figure is $1 \times 10^7$. The dashed lines indicate the average values for each QNU. On the left panel, the average excess noise for the four QNUs is 0.0542, 0.0525, 0.0527, and 0.0570 SNU in the $X$ quadrature, and 0.0507, 0.0505, 0.0490, and 0.0549 SNU in the $P$ quadrature, respectively. On the right panel, the average SKR values for the four QNUs are 20.73, 18.56, 20.41, and 17.43 Mbps in trusted condition. Despite the elevated excess noise in Bob 4 leading to a reduced key rate, all four QNUs maintain secret key rates above 15 Mbps, confirming the acceptable overall performance of the PSP-QPON. While in the untrusted condition, it has been proved that the PTMP protocol hits the theoretical key rate upper limit due to a tightly estimation of Eve’s accessible information \cite{tight_2026}.

We compare our work with the outstanding CV access network experiments conducted in recent years, and the result is shown in~Table~\ref{tab: SOA}. The average asymptotic SKR of our work is 19.29 Mbps over a 20 km equivalent fiber channel, where the equivalent channel consists of a 5 km fiber channel and a 3 dB attenuated free-space channel, converted to an equivalent fiber channel based on a fiber attenuation of 0.2 dB/km. Compared to other works (22.19 kbps@12.3 km\cite{Huang:20}, 1010 kbps@21 km\cite{Qi:24}, 549.2 kbps@11 km\cite{QAN_AA_2024}, 12050 kbps@15 km\cite{QAN_16node}), due to the advantages of the PSP protocol provides a higher equivalent repetition frequency, the average asymptotic SKR of our work is significantly higher than others.

\begin{table*}[htbp]
    \centering
    \caption{\textbf{Comparison of CV access network experiments.}}
    \label{tab: SOA}
    \begin{tabular}{cccccc}
        \hline
         Reference & Network type & QNUs/Capacity & Distance (km) & SKR (kbps) & Year\\
         \hline
         \cite{Huang:20} & Upstream & 2/2 & 12.3 & 22.19 & 2020\\
         \cite{Qi:24} & Downstream & 4/4 & 21 & 1010 & 2024\\
         \cite{QAN_AA_2024} & Downstream & 8/8 & 11 & 549.2 & 2024\\
         \cite{QAN_16node} & Downstream & 8/8 & 15 & 12050 & 2025\\
         Our work & Downstream & 4/4 & 20 & 19290 & 2025\\
        \hline
    \end{tabular}
\end{table*}

\begin{figure*}[htbp]
        \centering
        \includegraphics[width=0.8\linewidth]{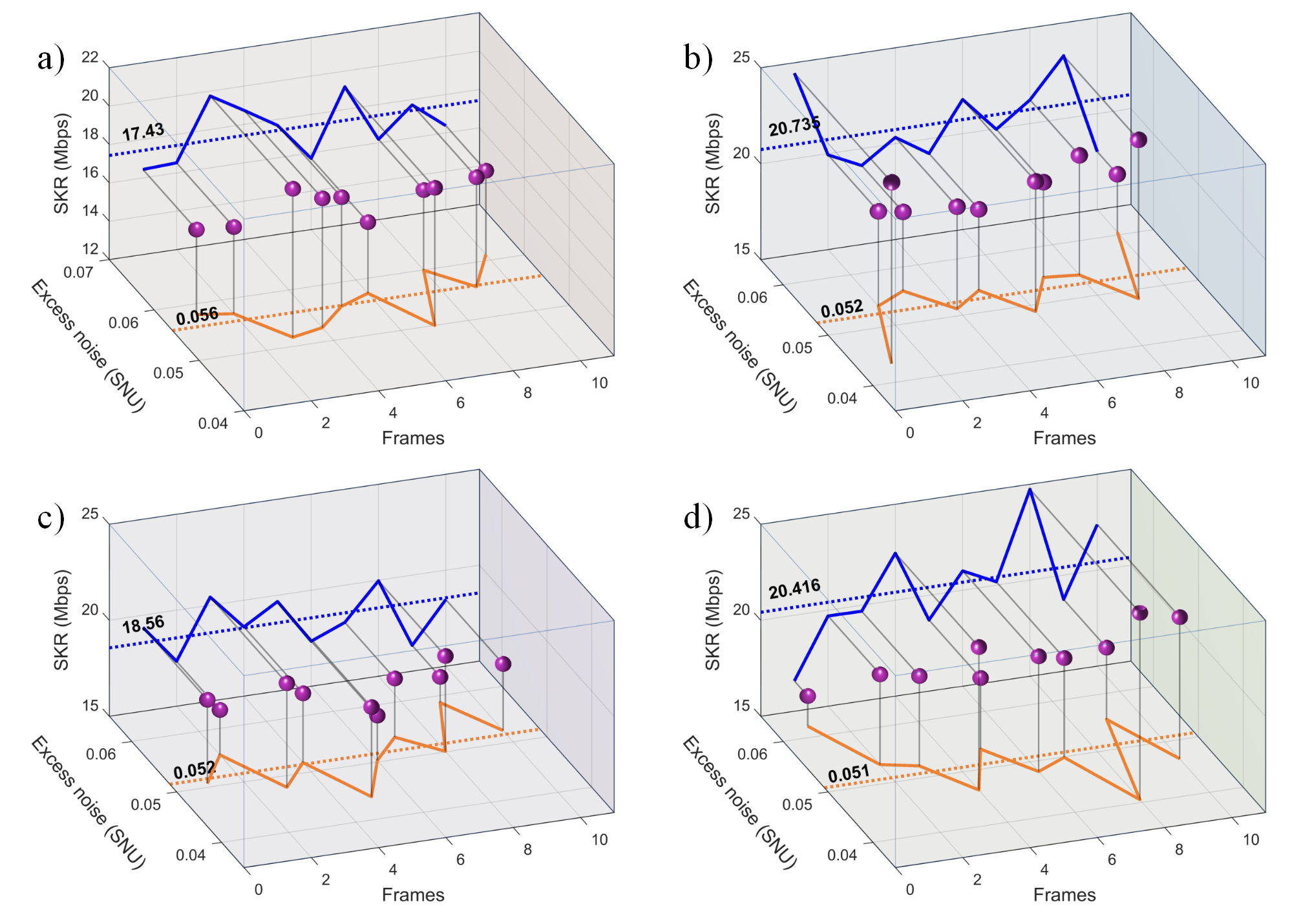}
        \caption{\textbf{The experimental results of PSP-QPON for 4 Bobs.} The SKR and excess noise of four QNUs. Each data point is calculated based on 25 sets of data frames, each 4e5 in length, forming a 1e7 data frame.}
        \label{fig: skr}
\end{figure*}

\begin{table*}[htbp]
        \centering
        \small
        \caption{\textbf{Key experimental parameters of PSP-QPON.}  $\beta$, reconciliation algorithm efficiency; $T$, the transmittance of the hybrid channel; $v_\text{el}$, the electrical noise of Bob; $SNR$, the signal noise ratio (SNR); $V_\text{A}$, equivalent modulation variance;  $\eta$, the detection efficiency; $f$, the equivalent repetition rate.}
        \label{tab: para}
        \begin{tabular}{c|ccccccc}
        \hline
        \textbf{Parameter} & $\beta$ & $\eta$ & $v_{el}$ (SNU) & $V_A$ (SUN) & $T$ (dB) & $SNR$ & $f$ (GHz) \\
        \hline
        \textbf{Value} & 0.96 & 0.56 & 0.12 & $4.28$ & $T_{Bob_1} = -10.77$ & $SNR_{Bob_1} = 0.112$ & 4\\
        &  &  &  &  & $T_{Bob_2} = -11.21$ & $SNR_{Bob_2} = 0.097$ & \\
        &  &  &  &  & $T_{Bob_3} = -10.94$ & $SNR_{Bob_3} = 0.107$ & \\
        &  &  &  &  & $T_{Bob_4} = -11.10$ & $SNR_{Bob_4} = 0.095$ & \\
        \hline
        \end{tabular}
\end{table*}

\section*{Discussion}

In this work, we propose and experimentally demonstrate a downstream fully passive quantum access network protocol with a free-space single-mode fiber hybrid channel, and the final key generation rate is up to 19.48 Mbps per QNU. The passive-state-preparation quantum passive optical network (PSP-QPON) extends the scope of PSP-CVQKD from PTP to PTMP networks, which provides a high key rate, high system stability, and low resource consumption system. Moreover, the network channel in this experiment is fully compatible with access networks in classical optical communications, which allows integration with existing optical infrastructure without the need for additional modifications, giving a perfect solution for LAN-wide quantum access network at home or a mobile terminal.

The practical deployment of PSP-QPON still faces several critical challenges, which could be further improved in the following studies. Firstly, since the ICR signal path is not polarization-maintaining, a PC needs to be added at both the transmitter and receiver ends to improve the detector efficiency, which introduces additional attenuation, reducing the system performance. Therefore, the system performance can be further improved by using polarization-maintaining ICR or designing a polarization diversity reception algorithm. Secondly, since the use of EDFA brings additional consumption at the receiver side, the use of high-intensity LO and the improvement of extinction ratio can be considered further to simplify the optical path at QNU. And developing the local LO scheme can further reduce the excess noise introduced by LO leakage and improve system performance while reducing the actual security vulnerabilities of the system. Furthermore, the QLT is located at a certain distance from the QNU. Therefore, for the PSP protocol that requires simultaneous detection at both ends, a time synchronization solution must be designed. Moreover, the higher the detection bandwidth and the smaller the memory of the acquisition device, the greater the demand for time synchronization accuracy. Fortunately, these issues can be resolved by using an acquisition board combined with high-precision GPS timing.

\section*{Methods}
\subsection*{Real-time shot noise acquisition}
Shot noise serves as the fundamental normalization unit in QKD systems, as it quantifies the amplification scale of the LO, which is crucial for evaluating the security bounds of key generation. However, shot noise demonstrates real-time variations primarily induced by laser intensity. Environmental perturbations, including temperature drifts, slight misalignments of optical components, and variations in channel loss, can also introduce non-negligible deviations in shot noise levels. These real-time variations may result in inaccurate normalization and practical attacks on QKD, thereby degrading the performance of QKD systems and compromising the security of the generated keys.

In this experiment, the real-time acquisition of shot noise is implemented by utilizing an optical switch to alternately switch the optical path between the signal measurement channel and the shot noise calibration channel. Specifically, the optical switch on the Bob side periodically diverts a portion of the optical signal to the ICR, where shot noise measurements are performed synchronously with the signal detection. To obtain precise real-time shot noise values, we alternately acquire shot noise and data, and subsequently the real-time shot noise value is computed using multiple sets of consecutive shot noise near the target data frame. This method effectively suppresses statistical errors in single measurements and ensures that the acquired shot noise accurately quantifies the LO intensity corresponding to the current data frame.

Real-time shot noise acquisition uses an optical switch at Bob’s side to alternately route the optical signal between the signal measurement and shot noise calibration channels. Specifically, the switch periodically diverts part of the signal to the ICR for synchronous shot noise measurement and signal detection. Shot noise and data are acquired alternately, and the real-time shot noise value is calculated from multiple consecutive shot noise datasets near the target data frame. This method suppresses single-measurement statistical errors and ensures the acquired shot noise accurately quantifies the current data frame’s LO intensity.

\subsection*{Frame synchronization and phase compensation}
In conventional digital signal processing (DSP) methodologies for CVQKD\cite{dsp1,dsp2}, the utilisation of pre-inserted training frames is a customary practice to ensure frame synchronisation and phase compensation between the two parties. This approach, however, introduces an additional overhead burden to the system. In order to address this limitation, a modulation-free phase compensation algorithm integrated with a gated recurrent unit (GRU) and local adaptive attention mechanism is proposed. This is paired with a fine-grained frame synchronization approach for the valid acquisition of raw data. Initially, Alice transmits a segment of her measurement data to Bob via a classical channel. Subsequently, Bob performs sliding cross-correlation analysis between the received data and his local measurement results to generate cross-correlation sequences. The validity of synchronisation is verified by comparing the maximum value of these sequences with a pre-calibrated threshold. In the event that the maximum value exceeds the threshold, frame synchronisation is confirmed, and the synchronised data sequence is extracted for subsequent phase compensation.

The core innovation of the proposed phase compensation algorithm lies in integrating alocal adaptive attention mechanism into the GRU-based temporal modeling framework. Unlike the traditional GRU, which serially processes data and equally weights all historical information, the proposed scheme focuses on the local temporal region with significant phase changes, thereby achieving the balanced optimization of "global temporal dependency modeling" and "key information enhancement". Specifically, the GRU unit is employed to capture the overall temporal correlation of phase shifts. At the same time, the local adaptive attention module is used to dynamically weight the phase data within a limited sliding window, thereby enhancing the model's perception of phase mutation points.

The training of models is based on phase drift data, which is compatible with the original workflow. In the context of the aforementioned experiment, Alice and Bob were tasked with the generation of phase data slices. To this end, they utilised synchronised sequences, employing a sliding window as a means to this end. For each slice, the local adaptive attention module calculates phase change rates to determine the size of the window and the attention weights. These are then fed into a GRU to predict the phase and compensate for it. In the training phase, 70\% of the data slices are utilised as the training set to collectively optimise the parameters of the GRU and attention module, while the remaining 30\% are allocated to the test set. In comparison with traditional GRU, the optimisation of attention weights has been demonstrated to enhance training efficiency and prediction accuracy. In the prediction stage, Alice and Bob obtain synchronised data, divide it into small segments, and Bob reveals half the data to assist in calculating the initial phase shift sequence. The model employs local adaptive attention to weight the revealed data, and GRU predicts the phase compensation amount for unrevealed data, thus avoiding the overfitting that is typically caused by traditional GRUs' equal treatment of all data.

%%%%%%%%%%%%%%%% MAIN TEXT FIGURES %%%%%%%%%%%%%%%

%%%%%%%%%%%%%%%% MAIN TEXT TABLES %%%%%%%%%%%%%%%

%%%%%%%%%%%%%%%% REFERENCES %%%%%%%%%%%%%%%

\clearpage % Clear all remaining figures and tables then start a new page

% The list of references goes after the main text and before the acknowledgements
% When preparing an initial submission, we recommend you use BibTeX, like this:
%
\bibliography{main} % for a file named science_template.bib
\bibliographystyle{sciencemag}

% After the paper has completed peer review and been revised ready for acceptance,
% you should comment out the lines above and copy-paste the contents of your .bbl
% file here instead. This will help ensure that our conversion software works correctly.
% Remember to re-run BibTeX first - check the timestamp!
%
% Example of the first three entries copy-pasted from science_template.bbl:
%
%\begin{thebibliography}{1}
%
%\bibitem{example}
%A.~N. {Author}, An example reference. \emph{Journal of Improbable Research}
%  \textbf{1}, 67 (2020).
%
%\bibitem{example2}
%F.~M. {Surname}, S.~{Author}, A second example. \emph{Interesting Research
%  Letters} \textbf{32}, 897 (2019).
%
%\bibitem{example_preprint}
%P.~{One}, P.~{Two}, P.~{Three}, {An unpublished preprint}. \emph{preprint}
%  (2021), arXiv:2101.12345.    
%
%\end{thebibliography}

%%%%%%%%%%%%%%%% ACKNOWLEDGEMENTS %%%%%%%%%%%%%%%

\section*{Acknowledgments}
We thank our colleagues for their contributions to the work cited.
\paragraph*{Funding:}
This work was supported by the Quantum Science and Technology-National Science and Technology Major Project (Grant No. 2021ZD0300703), Shanghai Municipal Science and Technology Major Project (2019SHZDZX01), the Key R\&D Program of Guangdong province (Grant No. 2020B0303040002), and the National Natural Science Foundation of China (No. 62101320).
\paragraph*{Author contributions:}
G.Z. conceived the research project. P.H. and H.Y. designed the scheme with assistance from T.W. and B.Z.. H.Y. and B.Z. carried out the experiments. X.J. and Y.X. contributed the approach to post-processing. H.Y. and P.H. wrote the manuscript with contributions from all authors.
\paragraph*{Competing interests:}
There are no competing interests to declare.
\paragraph*{Data and materials availability:}
All data needed to evaluate the conclusions in the paper are present in the paper and/or the materials cited herein. Additional data related to this paper may be requested from the authors.

\end{document}